\documentclass[iop,letterpaper,12pt]{emulateapj}
\usepackage{color}
\usepackage{dashrule}
\usepackage{multirow,bigstrut,bigdelim}
\usepackage{epsfig}
\def \bibpth {/Users/pveres/Dropbox/sync/bib/} 
\begin{document}
\title{Prospects for GeV-TeV detection of \\short gamma-ray bursts with extended emission}
\author{P. Veres\altaffilmark{1,*},  
P. \Mesz\altaffilmark{1}}
\altaffiltext{1}{Department of Astronomy and Astrophysics; Department of Physics;
Center for Particle and Gravitational Astrophysics; Institute for Gravitation and the Cosmos;
Pennsylvania State University, 525 Davey Lab, University Park, PA 16802, USA}
\altaffiltext{*}{{ Email: veres@gwu.edu, currently at: Dept. of Physics,
George Washington University, 725 21st st. NW, 20052 Washington DC.}}
\date{\today} 
\def\ve{\varepsilon}
\def\gbm{{\it GBM }}
\def\lat{{\it LAT }}
\def\fermi{{\it Fermi }}
\def\swift{{\it Swift }}
\newcommand{\Mesz}{{M\'esz\'aros}}
\def\mathnew{\mathsurround=0pt}
\def\simov#1#2{\lower .5pt\vbox{\baselineskip0pt \lineskip-.5pt
      \ialign{$\mathnew#1\hfil##\hfil$\crcr#2\crcr\sim\crcr}}}
\def\simg{\mathrel{\mathpalette\simov >}}
\def\siml{\mathrel{\mathpalette\simov <}}
\def\beq{\begin{equation}}
\def\enq{\end{equation}}
\def\bea{\begin{eqnarray}}
\def\ena{\end{eqnarray}}
\def\bitm{\bibitem}
\def\msun{M_\odot}
\def\L54{L_{54}}
\def\E55{E_{55}}
\def\et3{\eta_3}
\def\th1{\theta_{-1}}
\def\r07{r_{0,7}}
\def\x05{x_{0.5}}
\def\et600{\eta_{600}}
\def\et3{\eta_3}
\def\rph{r_{ph}}
\def\vareps{\varepsilon}
\def\fflunit{\hbox{~erg cm}^{-2}~\hbox{s}^{-1}}
\def\eps{\epsilon}
\def\ve{\varepsilon}
\def\muh{\hat{\mu}}
\def\cm{\hbox{~cm}}
\def\kpc{\hbox{~kpc}}
\def\s{\hbox{~s}}
\def\gev{\hbox{~GeV}}
\def\Jy{\hbox{~Jy}}
\def\TeV{\hbox{~TeV}}
\def\GeV{\hbox{~GeV}}
\def\MeV{\hbox{~MeV}}
\def\kev{\hbox{~keV}}
\def\keV{\hbox{~keV}}
\def\eV{\hbox{~eV}}
\def\G{\hbox{~G}}
\def\erg{\hbox{~erg}}
\def\s{{\hbox{~s}}
\def\cm2{\hbox{~cm}^2}}
\def\para{\parallel}
\def\Fl{\mathcal{F}}
\defcitealias{Veres+12magnetic}{VM12}
\defcitealias{Sari+00refresh}{SM00}
\begin{abstract}
We discuss the GeV to TeV photon emission of gamma-ray bursts (GRBs) within the
refreshed shock and the continuous injection scenarios, motivated by the
observation of extended emission in a substantial fraction of short GRBs.  In
the first model we assume that the central engine emits promptly material with
a range of Lorentz factors. When the fastest shell starts to decelerate, it
drives a forward shock into the ambient medium and a reverse shock in the
ejecta.  These shocks are reenergized by the slower and later arriving
material.  In the second model we assume that there is a continued ejection of
material over an extended time, and the continuously arriving new  material
keeps reenergizing the shocks formed by the preceding shells of ejecta.  We
calculate the synchrotron and synchrotron self-Compton radiation components for
the forward and reverse  shocks and find that prospective and current GeV to
TeV range instruments such as CTA, HAWC, VERITAS{ , MAGIC} and HESS have a
good chance to detect afterglows of short bursts with extended emission,
assuming a reasonable response time.  \end{abstract}

\section{Introduction}
\label{sec:intro}

Lightcurves of GRBs at GeV energies are becoming more numerous
\citep{Fermi+13grbcat} thanks to the observations of the Large Area Telescope
aboard the \fermi satellite \citep{Atwood+09LAT}. While most of the lightcurves
at GeV are for long GRBs, a fraction of short bursts also have a GeV component
(e.g. GRB 081024B \citet{Abdo+10-081024b} and GRB 090510
\citet{DePasquale+09-090510ag}).

The highest energy LAT photons associated with GRBs have $\sim 100 \GeV$ energy
in the comoving frame  \citep{Atwood+13newLAT}.  VERITAS observed the location
of 16 \swift detected bursts above $\sim 200 \GeV$ and found no associated
emission \citep{Acciari+11VeritasGRB}.

One of the important results of the \swift  era \citep{gehrels04} is the
discovery of extended emission following some of the short bursts
\citep{gehrels06,norris06}. \citet{Bostanci+13BATSEEE} found extended emission
following short GRBs in the BATSE data. An extended tail of the prompt emission
may be one of the tell-tale signs of a merger event \citep{Zhang+09origin}.
The detection of such  extended prompt emission in a high fraction of short
bursts prompts us to discuss the effects of a continued energy injection into
the afterglow external shock, and to discuss its effects on the high energy
lightcurves.  The effects of such late-arriving ejecta into the external
shocks, whether due to early ejection of a range of Lorentz factors or a
continued outflow, are especially important for the detectability with future
TeV range observatories, as it flattens the lightcurves providing a higher flux
at late times than in the standard impulsive, single Lorentz factor case.

Compact binary merger events are the most promising candidates for
gravitational wave detection with the LIGO and VIRGO instruments \citep[e.g.
][]{Abadie+12GWLIGO}.  The working hypothesis for the short GRB origin is
currently the binary merger scenario \citep[e.g.][]{Lee+07short, nakar07,
Littenberg+13binary}.  Electromagnetic counterpart observations are very
important for localization of the gravitational wave source
\citep{Aasi+13GWlocalize}, for additional follow up \citep[e.g.
][]{Evans+12GWfollowup} and for constraining the physical nature of the
gravitational wave signals. Furthermore, short bursts by themselves are of
significant interest for TeV range detections, since their spectra are
generally harder then those of long bursts.

Previous theoretical analyses typically considered synchrotron or synchrotron
self-Compton (SSC) emission components from top-hat like energy injection
episodes when addressing the high energy lightcurves \citep{dermer00,
Granot+02Dabreaks, panaitescu05, fanpiran06b,Fan+08highe, Kumar+09LATexternal,
Kumar+10external, He+09ll}.

In this paper we discuss the physical processes associated with the refreshed
shock and continued ejection scenarios in short GRBs, and we calculate the
expected lightcurves in different energy bands.  We present illustrative cases,
and discuss the observational prospects for GeV and TeV range observatories
already in operation or in preparation.

\section{Model for GeV and higher energy radiation from refreshed shocks}
\label{sec:model}

The basic idea of refreshed shocks is that the central engine ejects material
with a distribution of Lorentz factors (LFs) \citep{Rees+98refresh}.  This
happens on a short timescale and it can be considered instantaneous.  The blob
with the highest LF starts decelerating first by accumulating interstellar
matter. Forward and reverse shocks form. The trailing part of the emitted blobs
will catch up with the decelerated material reenergizing the shocks. This  will
result in a longer lasting emission by the reverse shock than the usual
crossing time (the latter being valid in the case without these refreshed
shocks).

The effect of the refreshed shocks is a flattening of the afterglow lightcurve
\citep{Sari+00refresh}.  This mechanism is considered a leading candidate for
the X-ray plateau (or shallow decay) phase
\citep{Nousek:2006ag,Grupe+13plateau}.

We consider the adiabatic evolution of the blast wave \citep{Meszaros+97ag} and
derive the relation between the LF and radius from energy conservation. 
At early times the afterglow may be in the radiative regime \citep{meszaros98}. 
In this case the conservation of momentum will yield the correct relation between
$R$ or $\Gamma$ and $t$.  $R$ and $\Gamma$ are linked by the $R\propto t
\Gamma^{ 2}$ relation \citep[e.g.][]{Meszaros+97ag,Waxman97angular}.

In our model of refreshed shocks, we assume { that} a total energy $E_{\rm
t} = 10^{53}\erg~ E_{\rm t,53}$ is released on a short timescale in a flow,
having a power law energy distribution with LF: $ E(>\Gamma)=E_M (\Gamma /
\Gamma_M)^{-s+1}$, from a minimum ($\Gamma_m$) up to a maximum LF ($\Gamma_M$).
The relation between the  total injected energy and the energy imparted to the
material with the highest LF ($E_M$) is {
\begin{equation}
E_{\rm t} = E(>\Gamma_m) = E_M
(\Gamma_m/\Gamma_M)^{-s+1}.
\label{eq:etot}
\end{equation}}
Once we set the total energy injected by the central engine and the two
limiting LFs, the initial energy ($E_M$) can be calculated for a given $s$.  It
is also possible to fix the extremal LFs and the initial energy (e.g. from
observations) and consequently the total injected energy can be determined. A
reasonable lower limit of injection LF ($\Gamma_m$) is a few tens
\citep{Panaitescu+98multi} and reasonable values for $\Gamma_M$ are a few
hundreds.

We introduce the $g$ parameter to describe the density profile of the circumburst 
environment, $n(R)\propto R^{-g}$ \citep{meszaros98}. The usual cases of homogeneous
interstellar medium and wind density profile are $g=0$ and $g=2$. The bulk of
our illustrative examples are for  $g=0$.

The characteristic energy corresponding to material with $\Gamma_M$ { is}
$E_M$. We will determine physical parameters based on the these quantities at
the deceleration radius as reference values for the time evolving quantities.
{ As an example, for $s=2$}, the deceleration radius in an interstellar
medium of $n = 1~ n_0 \cm^{-3}$ number density is: {  $R_M=(3 E_M/8 \pi n
m_p c^2 \Gamma_M^2)^{1/3}= 4.6 \times 10^{16} \cm~ E_{t,54}^{1/3} n_0^{-1/3}
\Gamma_{M,2.5}^{-1} (\Gamma_m/40)^{1/3} $ and the corresponding timescale
$t_M=R_M/2\Gamma_M^2 c= 15 \s~ E_{t,54}^{1/3} n_0^{-1/3}
\Gamma_{M,2.5}^{-3}(\Gamma_m/40)^{1/3}  $. Here we have used Equation
\ref{eq:etot} to derive the $E_t$ dependencies.  }

The evolution of the physical quantities results from energy conservation
\citep{Sari+00refresh} in the adiabatic case:
\begin{eqnarray}
R&=&R_M (t/t_M)^{(s+1)/(7+s-2g)}\\
\Gamma&=&\Gamma_M (t/t_M)^{-(3-g)/(7+s-2g)}\label{eq:gam}
\end{eqnarray}
We do not address the radiative case here, but mention, that it can be
accounted for by introducing an additional term in the power law index in the
above expressions  \citep{meszaros98}. { The number density of the ejecta,
when the reverse shock has crossed (which is approximately at the deceleration
radius) is: $n_b = E_M / 4 \pi R_M^2 m_p c^3 \Gamma_M^2 T = 1.1 \times 10^6
\cm^{-3}~ E_{t,54}^{1/3} \Gamma_{M,2/5}^{-1} (\Gamma_m/40)^{1/3} n_0^{2/3} T_0
$}, evaluated  { also} for $s=2$, where $T$ is the duration of the burst.
The magnetic field will be, as usually assumed, some fraction $\epsilon_B$ of
the total energy: {  $B=(32 \pi n m_p c^2 \epsilon_{B}
\Gamma_M^2)^{1/2}\approx 12.3~ {\rm G} ~ n_0^{1/2} \Gamma_{M,2.5}
\epsilon_{B,-2}^{1/2} $. } For simplicity, we use the same value for
$\epsilon_B$ across the forward and reverse shock.

The magnetic field will vary in time as: $B\propto t^{-(6+sg-g)/2(7+s-2g)}$.
The reverse shock optical scattering depth at the reference time is: $\tau_{\rm
RS}= (\sigma_T N)/(4 \pi R_M^2)=(\sigma_T E_M)/(4 \pi R_M^2 m_p c^2 \Gamma_M)$.
The time dependence results from the change of individual components:
$\tau_{\rm RS} \propto t^{-(2-s+sg)/(7-2g+s)}$.  The optical depth of the
forward shock region is $\tau_{\rm FS}=\sigma_T N_e/4\pi R^2\propto R^3
R^{-g}/R^2\propto t^{(1+s)(1-g)/(7-2g+s)}$. 
The cooling random LF is defined by an electron whose synchrotron
loss timescale is of the order of the dynamical timescale: $\gamma_{c,f}\propto
(\Gamma B^2 t)^{-1} \propto t^{(2-s+gs)/(7-2g+s)}$. The forward shock
electrons' injection LF is: $\gamma_{m,f}\propto \Gamma  \propto
t^{(2-s+gs)/(7-2g+s)}$. The corresponding synchrotron frequencies:
$\ve_{m,f}\propto \Gamma \gamma^2_{m,f} B \propto t^{-(24-7g+gs)/(2(7-2g+s))}$.

{\it Synchrotron component - } 
Initially the forward shock will be in the fast cooling regime. The injection
electron LF is: $\gamma_m= \Gamma \epsilon_e m_p (p-2)/(p-1) m_e$, while the
cooling LF is $\gamma_c=6 \pi m_e c/{ (1+Y)}\sigma_T \Gamma_M B^2 t $, {
where Y is the Compton parameter \citep{Sari+01ic}}. The corresponding
synchrotron energies will be $\ve_{m,c}^{\rm syn}=\Gamma \gamma_{m,c}^{ 2} h
q_e B/2\pi m_e c $, where the $m$ and $c$ indices mark the injection and
cooling cases respectively. The spectral shape will be broken power law
segments with  breaks at these energies.  The peak flux of the synchrotron
results from considering the individual synchrotron power of the shocked
electrons: $F_{\ve,{\rm peak}}^{\rm syn} = ({ 4\pi n R^3 / 3})(m_e c^2
\sigma_T \Gamma B /3q_e) /(4\pi D_L^2) $. The maximum synchrotron energy is
derived from the condition that the electron acceleration time is the same
order as the radiation timescale, $\ve_{\rm MAX,syn}\approx 2.5 \GeV (\Gamma_M
/100)$ \citep{dejager+96maxsyn}. 

The reverse shock electrons cool at the same rate as the forward shock
electrons and their cooling LF ($\gamma_c$) will be the same. Their injection
LF however is lower by a factor of $\Gamma$, due to the larger density in the
ejecta.  The reverse shock cooling frequency will be the same as the forward
shock, while the injection frequency is lower by $\Gamma^2$. The peak of the
reverse shock flux is larger by a factor of $\Gamma$.

{\it Synchrotron self-Compton component - }
Synchrotron photons emitted by the forward and reverse shock electrons will act
as seed photons for inverse Compton scattering on their parent electrons.
Inverse Compton radiation is expected to occur both in the forward and in the
reverse shock \citep{Sari+01ic}.  For the SSC spectral shape we consider  a
simplified approximation, neglecting the logarithmic terms which induce a
curvature in the spectrum \citep{Sari+01ic, Gao+13ic}.  The SSC components will
have breaks at $\ve_{m,c}^{\rm SSC}\approx 2 \gamma_{m,c}^2 \ve_{m,c}^{\rm syn}
$.  The SSC peak flux $F_{F,R}^{\rm SSC}\approx \tau_{F,R} F_{F,R}^{\rm syn}$
for the forward (F) and the reverse (R) shocks respectively. 

{\it Other components - } There can be an inverse Compton upscattering between
the RS electrons and the FS photons, FS electrons and RS photons. We estimated
the peaks of these components and found them to be subdominant for { the
parameters presented here.}

{\it Effects suppressing the $\TeV$ radiation -}
\begin{itemize}
\item {\it Klein-Nishina suppression -} At high energies the inverse Compton
component can be affected by Klein-Nishina effects. These occur above
$\ve\approx\Gamma \gamma m_e c^2$ and cause a steepening in the spectrum
\citep{Guetta+03plerion, Nakar+09KN, Wang+11KN}.

\item {\it Pair absorption effects in the source-} In the external shock
region, the optical depth to $\gamma\gamma\rightarrow e^{\pm}$ process is
negligible for most of the parameter space. This is mostly due to the low flux
at TeV energies and large radiating volumes resulting in low compactness
parameters \citep{Panaitescu+13pairs}. 

\item {\it Pair suppression on the extragalactic background light (EBL) -} To
account for the effects of the pair creation with UV to infrared photons of the
extragalactic background light we use the model of \citet{Finke+10ebl}. This
suppression is relevant generally above $100 \GeV$, but depends  significantly
on  the source redshift. As an example, the lightcurve at $1\TeV$ is suppressed
by a factor of $\sim 1/3$ for a source at $z=0.1$.
\end{itemize}

{ Regarding the last point: the pairs created by TeV range source photons
can upscatter the cosmic microwave background (CMB) photons to GeV energies
\citep{plaga95}. This would result in a GeV range excess in the spectrum. To
account for this component, detailed numerical treatment is needed \citep[see
e.g.][]{Dermer+11magn} which is not the scope of this work. By neglecting this
component, we assume the intergalactic magnetic field (IGMF) is strong enough
to change the direction of the pairs by roughly the beam size of the observing
instrument (e.g.  few $\times 0.1 ^\circ$ for Fermi around $1\GeV$).  The
deflection angle is $\Delta \theta\approx (R_{\rm cool}/R_{\rm
Larmor})(\lambda_\gamma/D) $ where the expression consists of the cooling length
of pairs on the CMB photons, the Larmor radius of the pairs, the mean free path
of TeV photons on the EBL, and the source distance in this order. For a source
at $z=0.2$ the mean free path of a TeV photon is of the order of a few $\times
100$ Mpc, with uncertainties depending on the EBL model.  Assuming the
coherence length of the IGMF larger than the cooling length of the pairs, for
the magnetic field we get $B>6\times 10^{-16} \G (\ve_{\rm
source}/{\TeV})^{2}$.

On the other hand, such a magnetic field will result in an exceedingly long
time delay for the arrival of the GeV photons. Any magnetic field value in
excess of $\sim 8 \times 10^{-20} \G$ will give a delay longer than $10^5 \s$,
making the association of the excess GeV flux with the GRB difficult.

}

\section{Continued Injection Models}
\label{sec:continj}

Late energy injection in the forward shock can also occur if there is a
prolonged central engine activity following the energy injection episode
responsible for the prompt emission \citep{Blandford+76bm,dailu98b,
Zhang+01mag}. { While in the case of short bursts, in a binary neutron star
coalescence scenario the disrupted material is accreted on timescales of $10
{\rm ms}$ \citep{Lee+07short}, it is possible that the merger results in the
formation of a magnetar, which injects energy on a longer scale.  The magnetar}
injects energy as it spins down, but it can occur also for a black hole central
engine. An initial instantaneous energy injection can also be incorporated,
which dominates the initial behavior.  At sufficiently late times (of the order
of one to a few times the deceleration time), the effect of the continuous
injection dominates over the initial injection episode.  Using the energy
conservation and setting an power law injection profile, we can derive dynamic
quantities in a similar fashion to \citet{Sari+00refresh}, \citet{Zhang+01mag}
or Section \ref{sec:model}.

The conservation of the total energy $E_t$ at sufficiently late times  
\citep{Zhang+01mag} reads
\begin{equation} \label{eq:e}
E_t =\frac{L_0 t}{\kappa+q+1} \left(\frac{t}{t_0}\right)^q + E_{\rm inst.}
\left(\frac{t}{t_0}\right)^{-\kappa},
\end{equation}
where $q$ is power law temporal index for the injected luminosity and $\kappa$
governs the impulsive energy input ($E_{\rm inst.}$) at early times. 
$t_0$ marks  the start of the self similar phase and $\kappa+q+1>0$ has to be
fulfilled.

\section{Lightcurve Calculations}
\label{sec:lc}

\subsection{Refreshed shock model lightcurves}
\label{sec:refresh}

\begin{figure}[htbp]
\begin{center}
\includegraphics[width=0.899\columnwidth,angle=0]{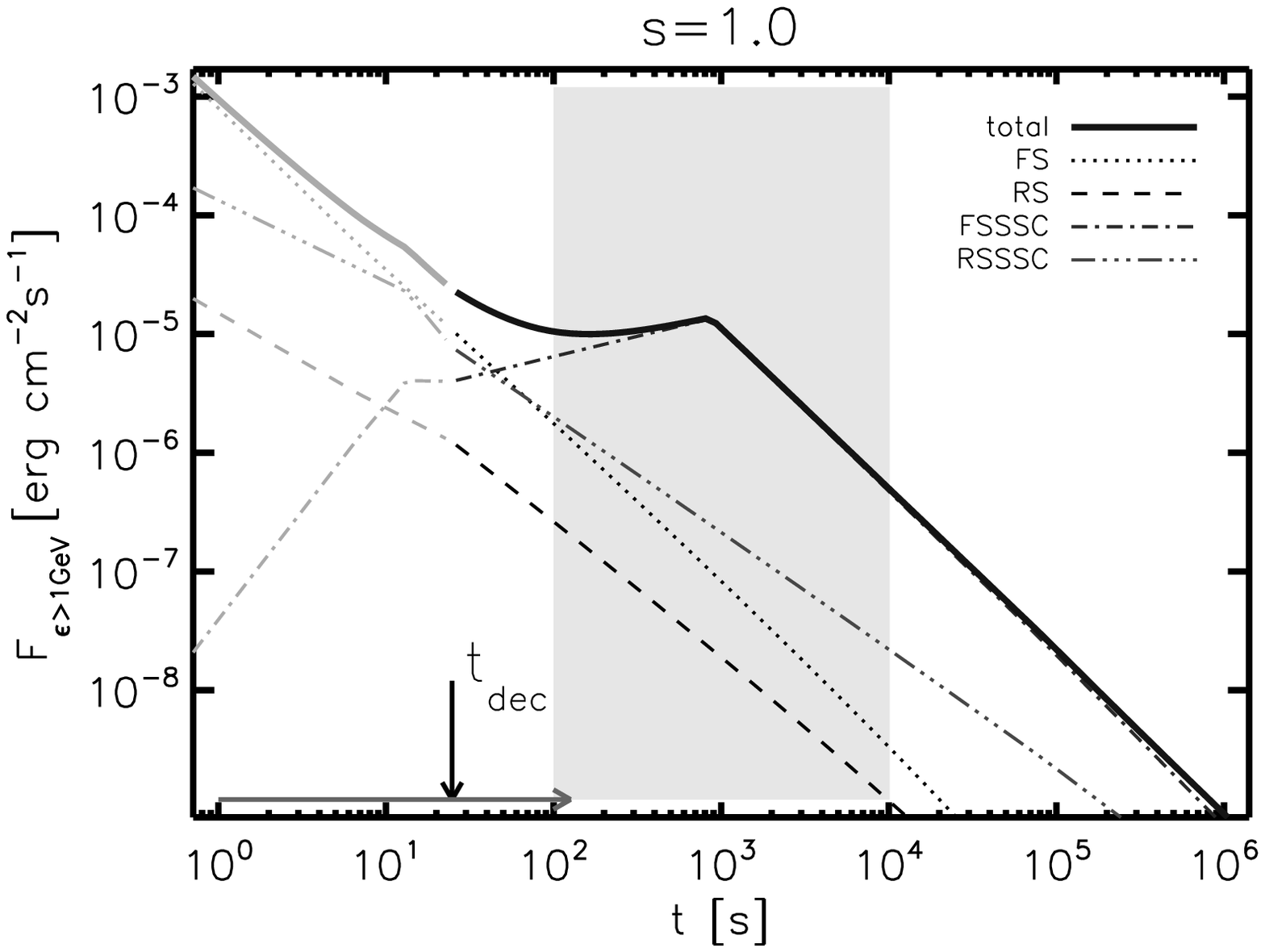}
\includegraphics[width=0.899\columnwidth,angle=0]{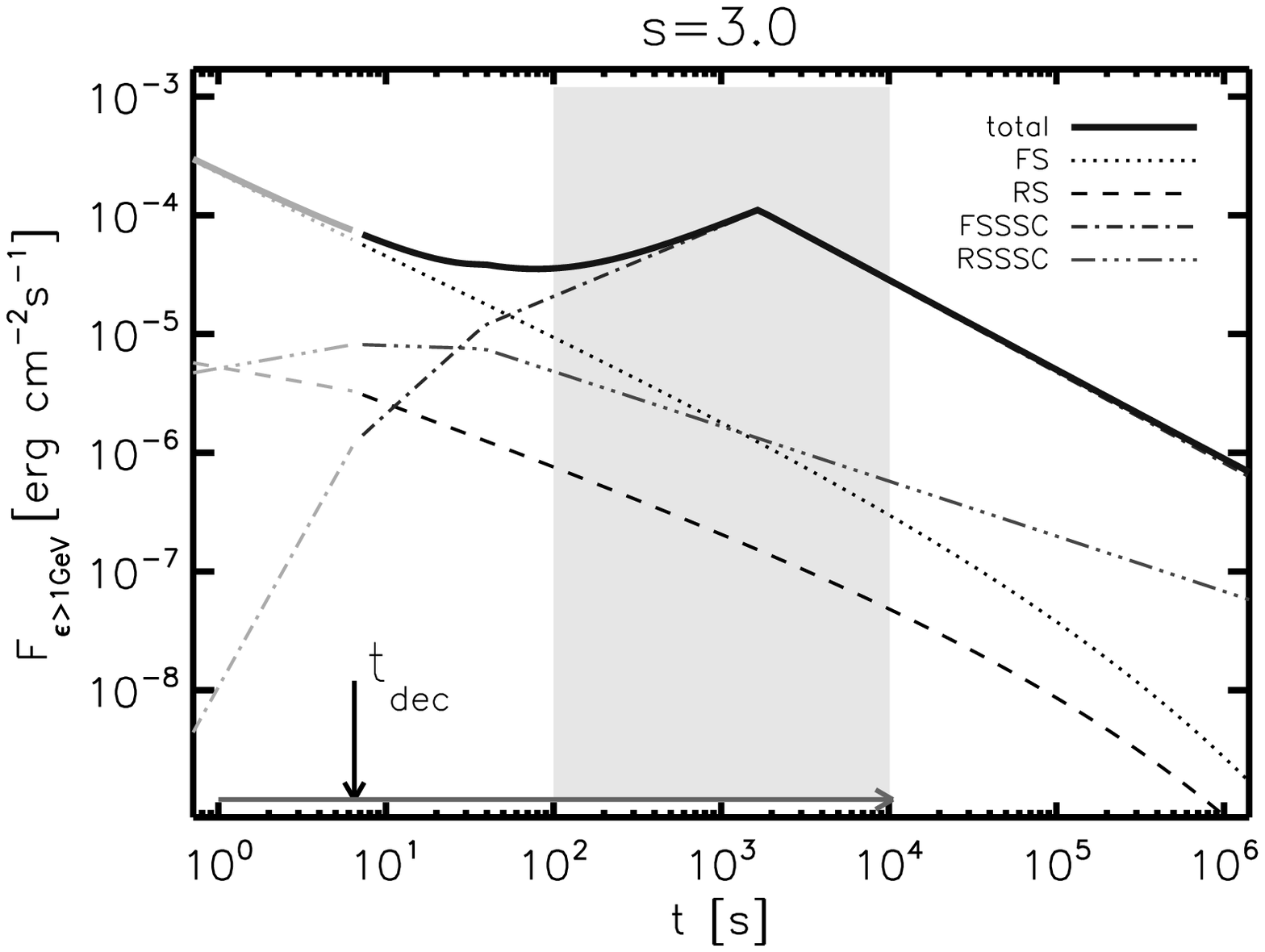}
\caption{Lightcurves showing individual components for $s=1$ and $s=3$ cases
above $1\GeV$. The horizontal arrow marks the duration of the fast cooling in
the forward shock, the vertical arrow marks the deceleration time of the
fastest shell. The lighter shades of the curves indicate flux values prior to
the deceleration time, where the Lorentz factor is constant. The shaded region
marks an optimistic observation window with Cherenkov telescopes. The notation:
FS is short for forward shock, RS for reverse shock, and SSC is synchrotron
self-Compton. } 
\label{fig:lc1} 
\end{center}
\end{figure}

For the refreshed shocks we adopt a set of realistic nominal parameters, which
are: $\Gamma_m/40 = \Gamma_M/300 = E_{t,53} = s/3 = n_{-1} = \epsilon_{e,-0.5}
= \epsilon_{B,-2} = p/2.5 = z/0.2=1$, and $g=0$.  Unless stated otherwise, we
use these parameters in the rest of the article.  Here, $\epsilon_{e}$ is the
fraction of the energy carried by electrons, $p$ is the exponent in the power
law distribution of the shocked electrons, and $z$ is the source redshift.  The
lightcurves are initially dominated by the forward shock synchrotron emission.
Generally, characteristic break frequencies will decrease in time.  Initially,
the GeV range is above the injection synchrotron frequency and the electrons
are in the fast cooling regime.  The slope is expected to be $-(-4 - 4 s + g +
s g + p/2 (24 - 7 g + s g)) / 2 ( 7 + s - 2g) = -0.7$ (see
\citet{Sari+00refresh} and Figure \ref{fig:lc1} (bottom) at early times). After
$\sim 10^2 \s$, or the order of the deceleration time, the forward shock SSC
emission starts to dominate the flux.  The GeV range is initially (while
$t<10^3 \s$) below the cooling and characteristic SSC frequencies.  The SSC
lightcurve has a break when $\ve_{\rm c,SSC}$ passes from above and reaches its
peak after $\ve_{\rm m,SSC}$ passes. By this time this is the dominating
component by one order of magnitude. The emergence of the SSC component results
in a plateau, then a decay with $(-4+8s-36(p-1)/2+(4-8s)/2)/(14+2s)=-0.85$
temporal index, calculated for the $\ve_{\rm obs} > \ve_{\rm m,SSC}, \ve_{\rm
c,SSC}$ case. The illustrative cases in the figures show a sharp break when
some characteristic frequency sweeps through the observing window. In reality
we expect a more gradual transition \citep{Granot+02Dabreaks}.

Temporal indices for a given ordering of $\ve_{\rm obs}, \ve_{\rm m,SSC} $ and
$ \ve_{\rm c,SSC}$ for one of the components can be found by considering the
instantaneous synchrotron (or SSC) spectrum with spectral indices of $\{1/3,
-1/2 $  or $-(p-1)/2, -p/2\}$ with increasing energy for fast or slow cooling
respectively.  Thus, for example, the reverse shock SSC with $\ve_m^{\rm RSSSC}
<\ve_{\rm obs} < \ve_m^{\rm RSSSC}$ will have a flux: $F_{\ve} (t)=F^{\rm
RSSSC}_{\ve,{\rm peak}} (\ve_{\rm obs}/\ve_{m}^{\rm RSSSC})^{-(p-1)/2} \propto
t^{-(18 - 3g - 12 s + 7 g s + p (30 - 6 g - 12 s + 8 g s)) / 6 (7 + s - 2 g)} =
t^{0.55}$.

The temporal slope of the overall lightcurve will depend on the slope of the
dominating component(s).  Similarly to  \citet{wang01a}, we also find that with
certain parameters, the RSSSC can dominate the lightcurve at early times. 

\begin{figure}[htbp]
\begin{center}
\includegraphics[width=0.899\columnwidth,angle=0]{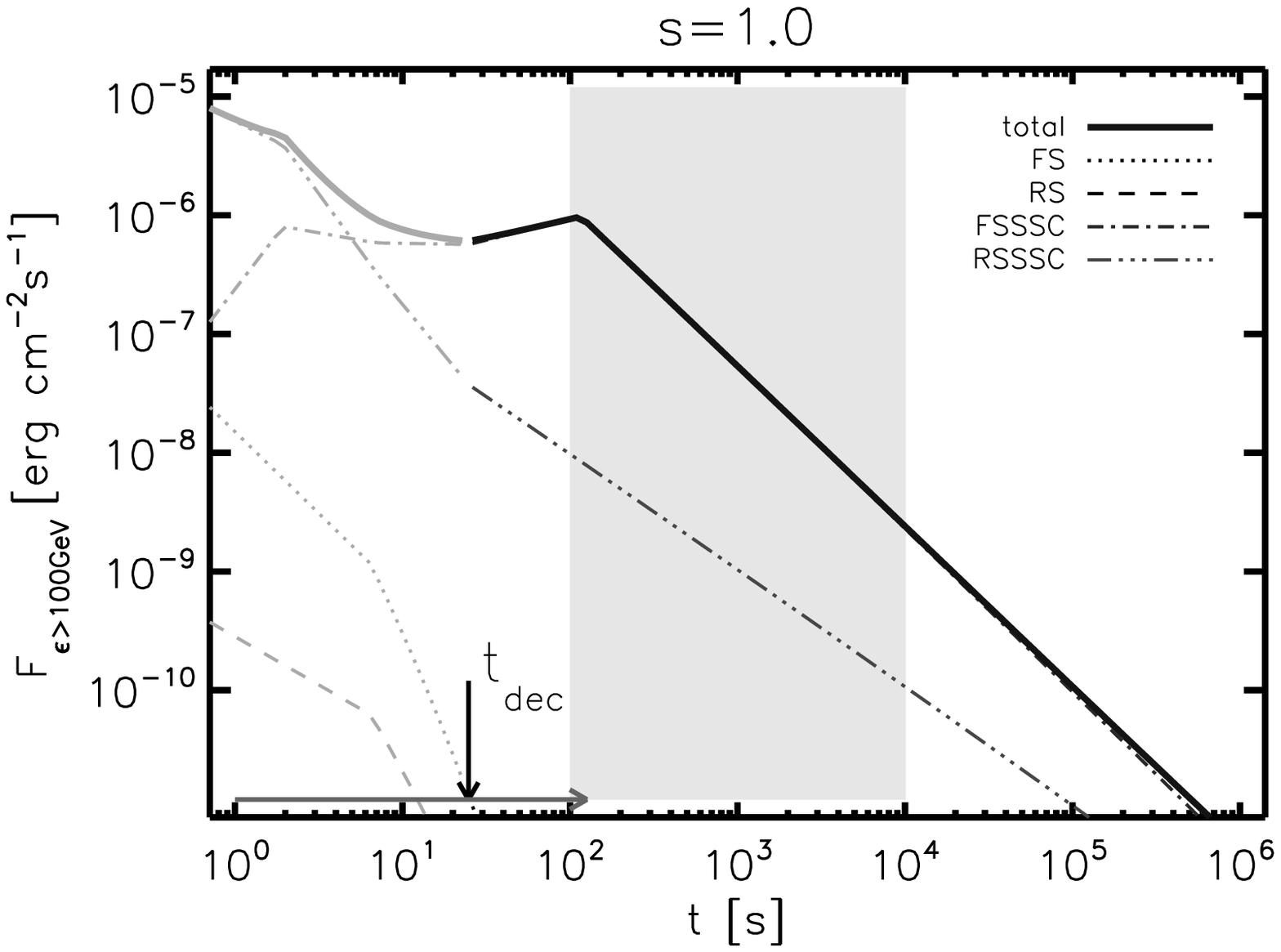}
\includegraphics[width=0.899\columnwidth,angle=0]{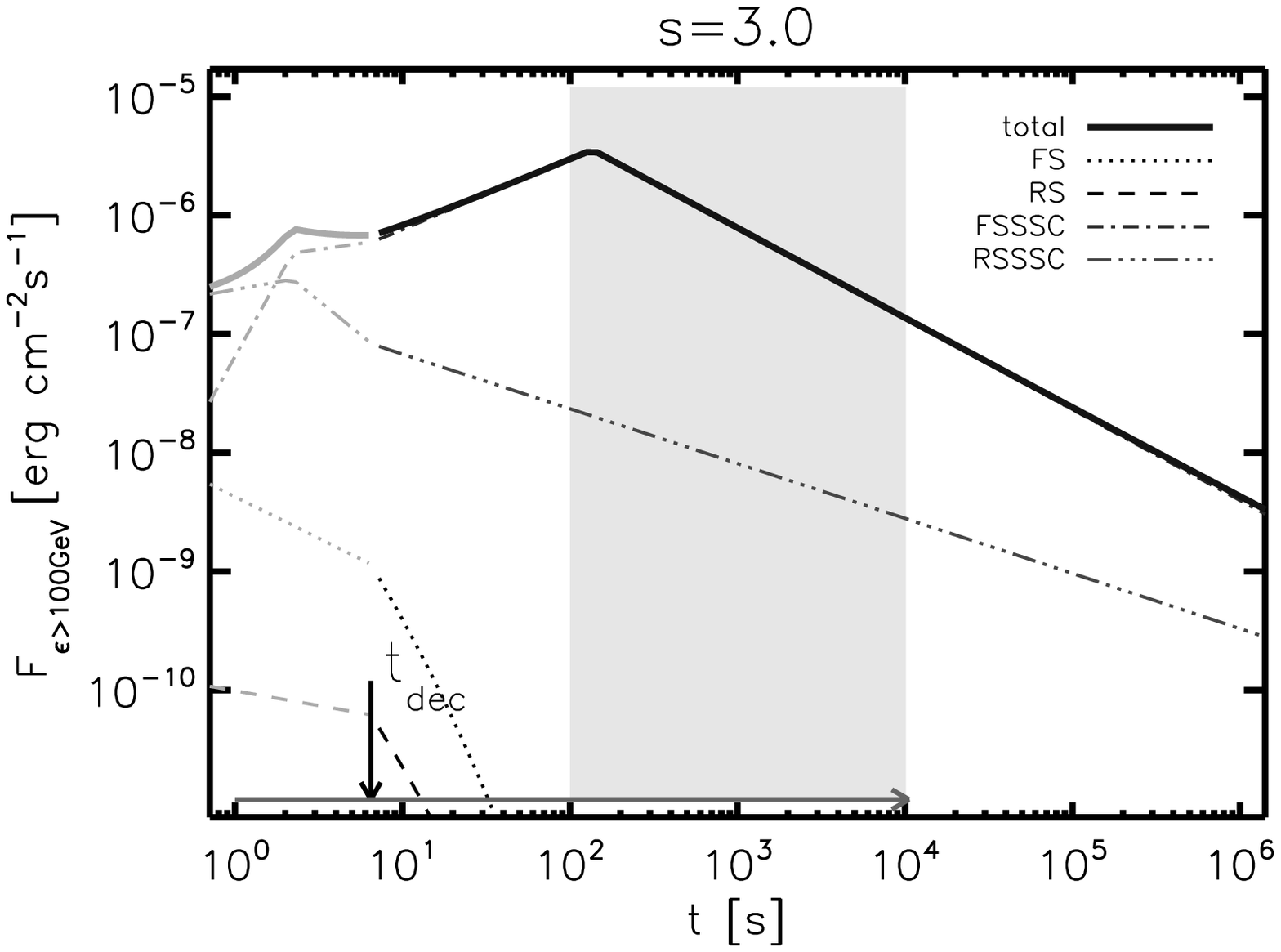}
\caption{Lightcurve at $100 \GeV$. Notations are the same as on figure
\ref{fig:lc1}. } 
\label{fig:lc3} 
\end{center}
\end{figure}

\subsection{Effect of changing microphysical parameters}

Cases with density from  $10^{-2}$ to $10 \cm^{-3}$ have essentially the same
lightcurves at late times in both presented cases (see Figure \ref{fig:lcnext}
presenting the flux above $100 \GeV$). Lower values of the density results in
fainter fluxes. There is no significant effect on the peak time of the
lightcurve (or the end of the plateau phase in the no injection case) happening
around $\sim 100 \s $ after the burst. 

The effect of changing magnetic parameter $\epsilon_B$ has a more accentuated
effect. When other parameters remain unchanged the $\epsilon_B\approx 10^{-3}$
gives the largest flux both in the injection and the no-injection cases (see
Figure \ref{fig:lcepsB} Both lower and higher values generally give a lower
flux. The peak time of the lightcurve  correlates with the value of $\epsilon_B$:
for lower $\epsilon_B$ the peak is earlier.

\begin{figure}[htbp]
\begin{center}
\includegraphics[width=0.899\columnwidth,angle=0]{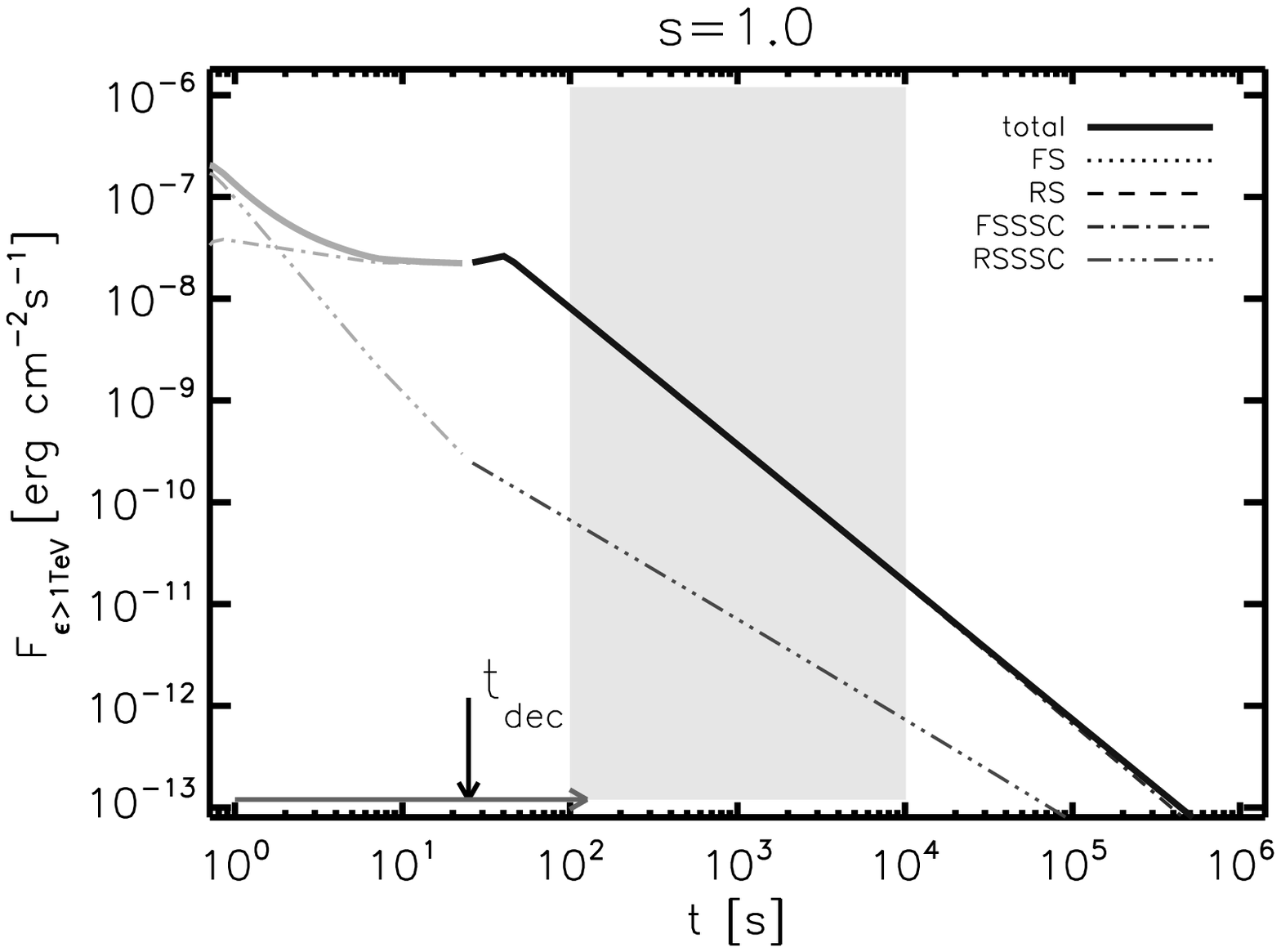}
\includegraphics[width=0.899\columnwidth,angle=0]{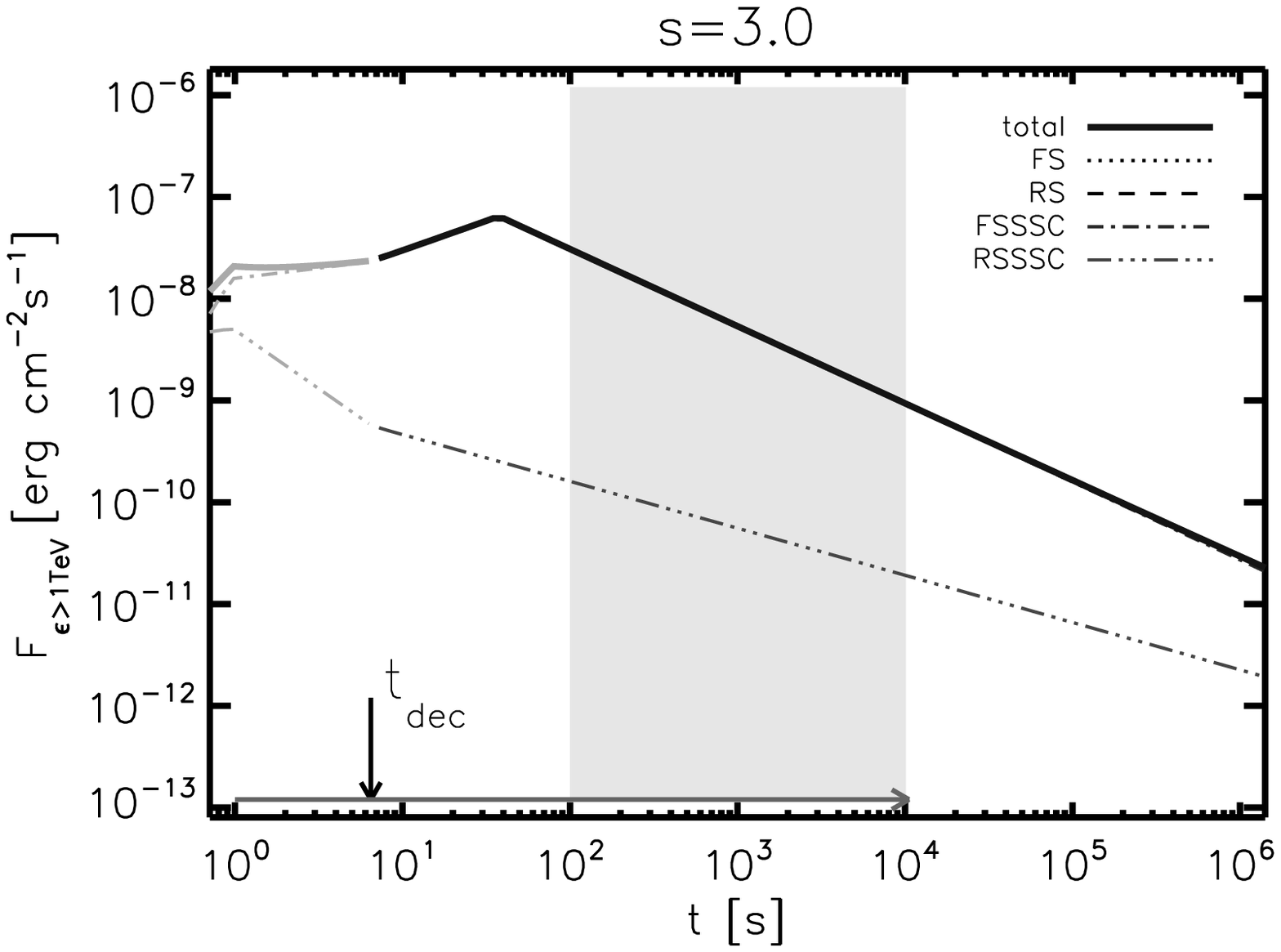}
\caption{Lightcurve at $1 \TeV$. Notations are the same as on figure
\ref{fig:lc1}. Note that the synchrotron components are insignificant due to
the cutoff at the maximum attainable synchrotron frequency.} 
\label{fig:lc4} 
\end{center}
\end{figure}

\begin{figure}[htbp]
\begin{center}
\includegraphics[width=0.899\columnwidth,angle=0]{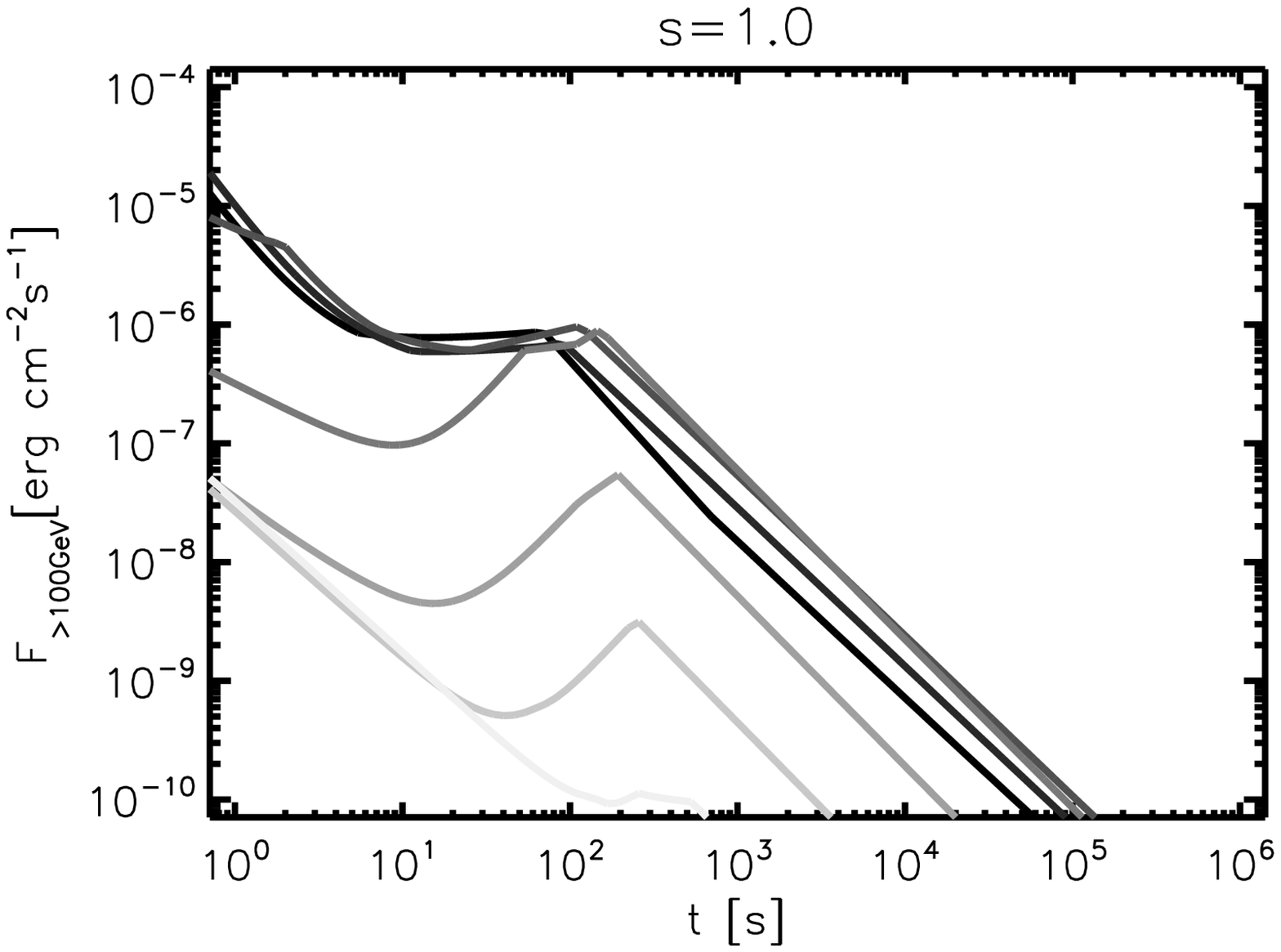}
\includegraphics[width=0.899\columnwidth,angle=0]{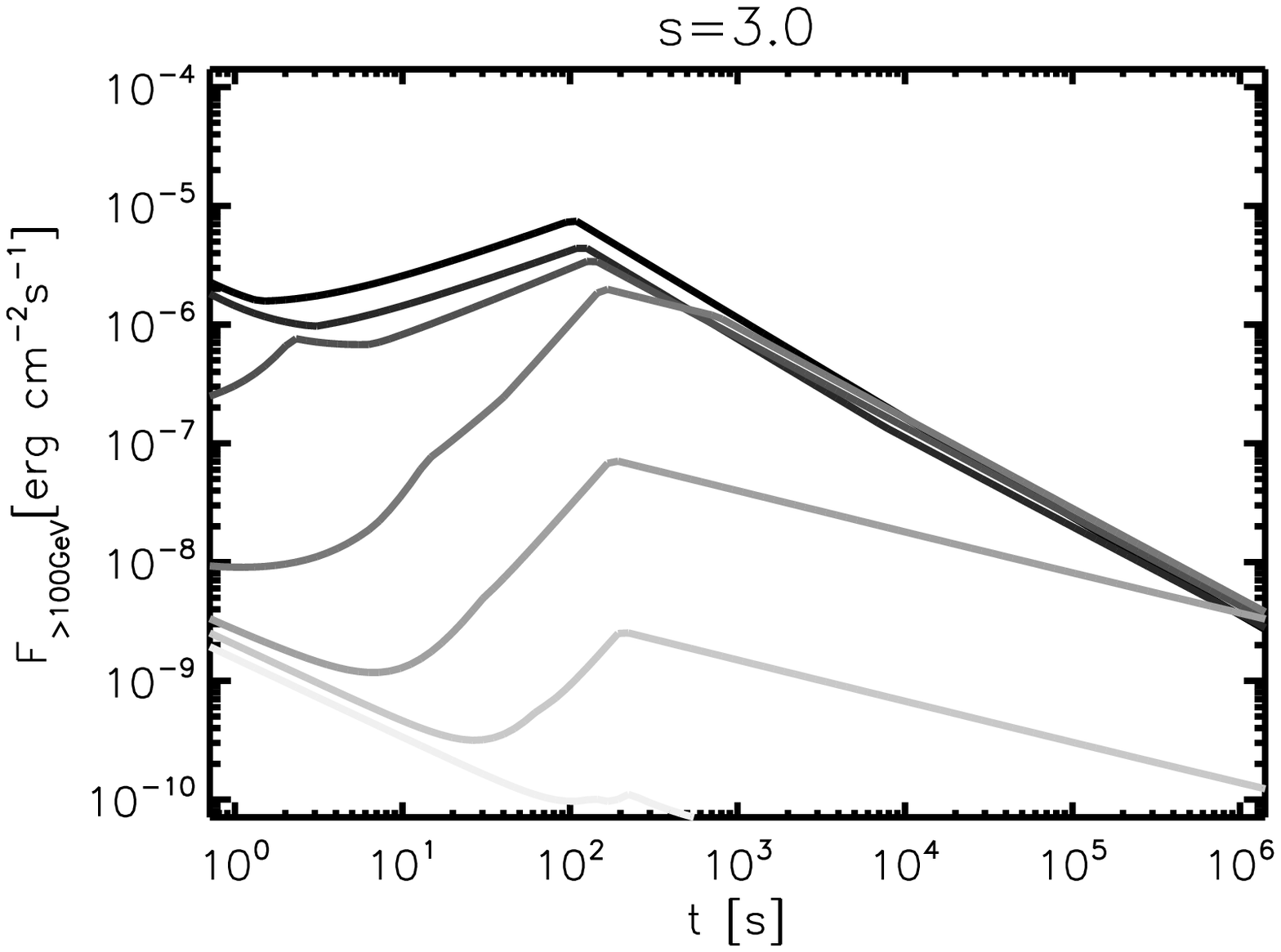}
\caption{Lightcurves at $100 \GeV$ showing the effect of a changing interstellar
density. The lightest curve is for $n_{\rm ext}=10^{-5} \cm^{-3}$, the
darkening shades represent a 10 fold increase, ending at $n_{\rm ext}=10
\cm^{-3}$. The top figure is for $s=1$, while the bottom figure is for $s=3$.}
\label{fig:lcnext} 
\end{center}
\end{figure}

\begin{figure}[htbp]
\begin{center}
\includegraphics[width=0.899\columnwidth,angle=0]{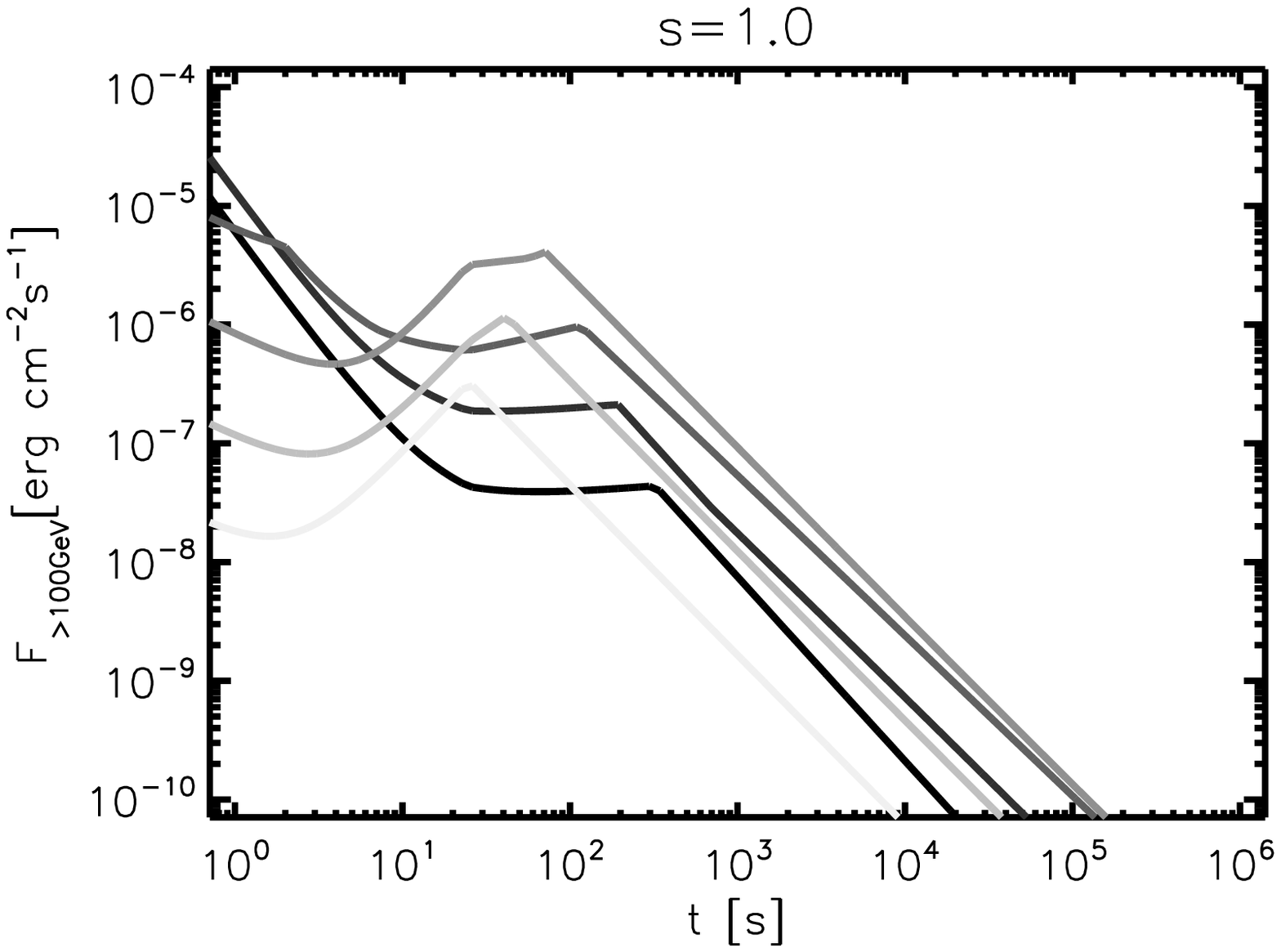}
\includegraphics[width=0.899\columnwidth,angle=0]{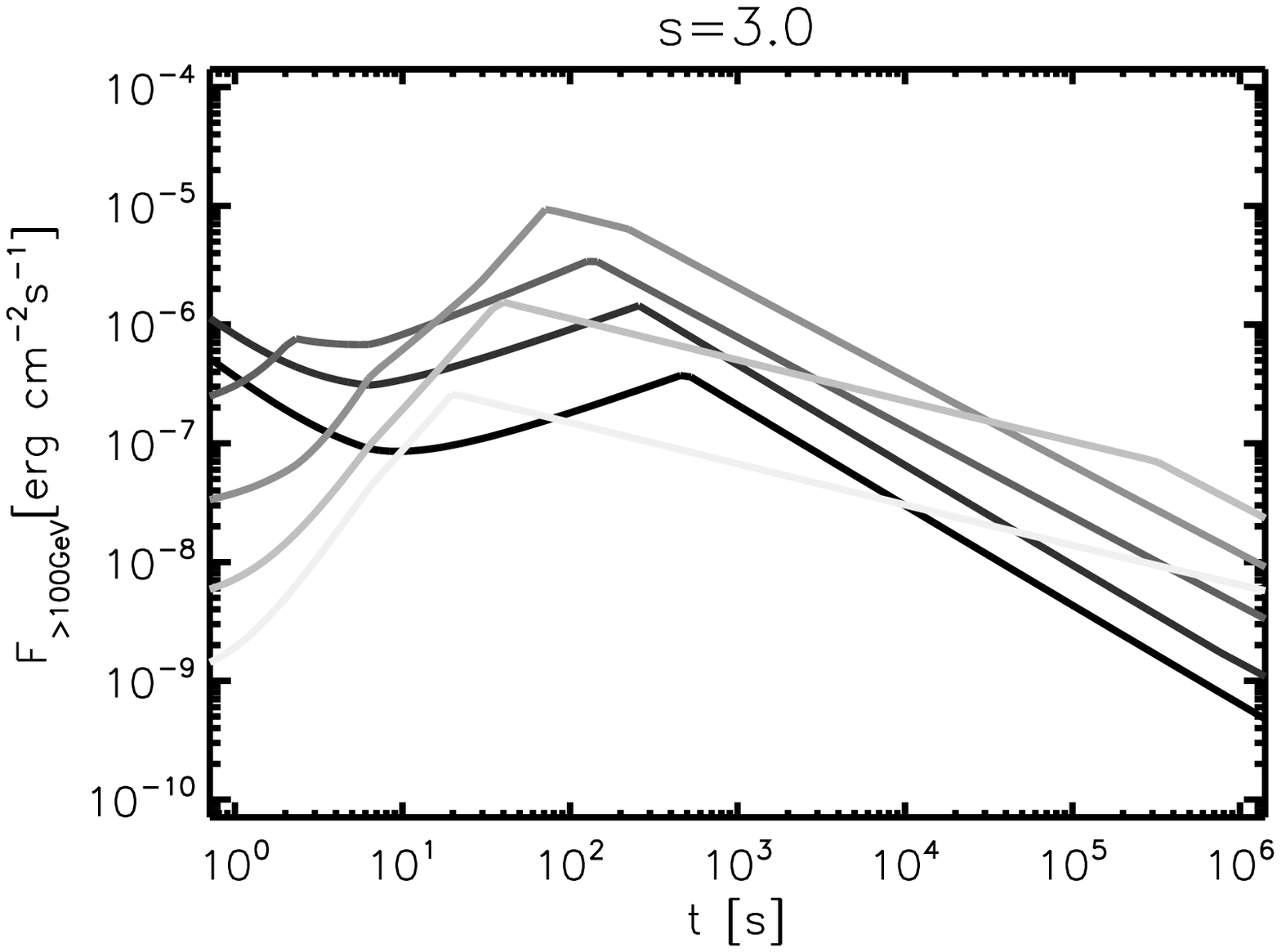}
\caption{Lightcurves at $100 \GeV$ showing the effect of a changing magnetic
parameter ($\epsilon_B$).  The lightest curve is for $\epsilon_B=10^{-5}$, the
darkening shades represent a tenfold increase, ending at $\epsilon_B=1$.}
\label{fig:lcepsB} 
\end{center}
\end{figure}

The lightcurves of afterglows arising from the refreshed shock scenario are
shown for three different energy ranges.  The first plot is for the range above
$1 \GeV$, since there is abundant data from LAT in this range, allowing a
straightforward comparison to past observations (Figure \ref{fig:lc1}).  The
next plot shows the lightcurves above $100 \GeV$, which represent  a transition
between the Fermi and Cherenkov telescope sensitivities for GRBs (Figure
\ref{fig:lc3}).  The third plot shows the lightcurves above $1 \TeV$, a so far
observationally fallow energy range, but one where the sensitivity of TeV range 
telescopes is currently being significantly increased, and where CTA and HAWC 
may bring breakthroughs in the near future (Figure \ref{fig:lc4}).

\begin{table}
\begin{center}
  \begin{tabular}{lccc} \hline\label{tab:tempidx}
	& $\ve_{m}^{\rm SSC}$	& $\ve_{c}^{\rm SSC}$	& $F^{\rm SSC}_{\ve_{\rm
	peak}}$ \\
\hline \hline
F	&$-\frac{36-11g+gs}{2(7+s-2g)}$		& $-\frac{-4+8s-3g-7gs}{2(7+s-2g)}$
&$-\frac{4-8s+g+5gs}{2(7+s-2g)}$	\\ 
\vspace{0.3cm}
R	&$-\frac{12-3g + gs}{2(7+s-2g)}$	& $-\frac{-4+8s-3g-7gs}{2(7+s-2g)}$
&$-\frac{16-8s-3g+5sg}{2(7+s-2g)}$\\
\hline
\end{tabular}
\caption{Temporal indices for calculating the self-Compton radiation behavior.}
\end{center}
\end{table}

\subsection{Flattening and peak of SSC component}
Previous studies predicted that an SSC component provides a flattening of the
lightcurve \citep{Sari+01ic,dermer00}. The  injection scenario makes this
flattening more pronounced or even results in an increase of flux.  The peak of
the SSC components can be found from: $\ve_{\rm obs}=\max\{\ve_m^{\rm
FSSSC},\ve_c^{\rm FSSSC}  \}$ for fast cooling of the forward shock. 

As an example, at $100 \GeV$, the peak occurs in the fast cooling regime, thus
we can calculate the peak time from $\ve_{\rm obs}=\ve_m^{\rm FSSSC}$.
Expressing the SSC injection frequency, we get $\ve_{\rm obs}\propto \Gamma^6
\eps_e^4 \epsilon_B^{1/2} g(p)^4 n_{\rm ext}^{1/2}$, where $g(p)=(p-2)/(p-1)$.
Taking the observing energy as a constant and keeping in mind that the time
dependence is carried by $\Gamma$ from Equation \ref{eq:gam} we find the peak
time varies as:
\begin{equation}
t_{\rm peak}\propto \eps_e^{2(7+s)/9} \epsilon_B^{(7+s)/36} g(p)^{2(7+s)/9}
n_{\rm ext}^{(s-5)/36}.
\end{equation}
This is in agreement with the change of the peak with $n_{\rm ext}$ and
$\epsilon_B$ presented in the previous subsection (figures \ref{fig:lcnext} and
\ref{fig:lcepsB}): $t_{\rm peak}$ has a very weak dependence on $n_{\rm ext}$
while a change of $10^5$ in $\epsilon_B$ results in a change of $\approx 13$
and $\approx 25$ in the peak time in Figure \ref{fig:lcepsB} (for $s=1$ and
$s=3$ respectively).

\subsection{Continued injection model lightcurves}

\begin{figure}[htbp]
\begin{center}
\includegraphics[width=0.899\columnwidth,angle=0]{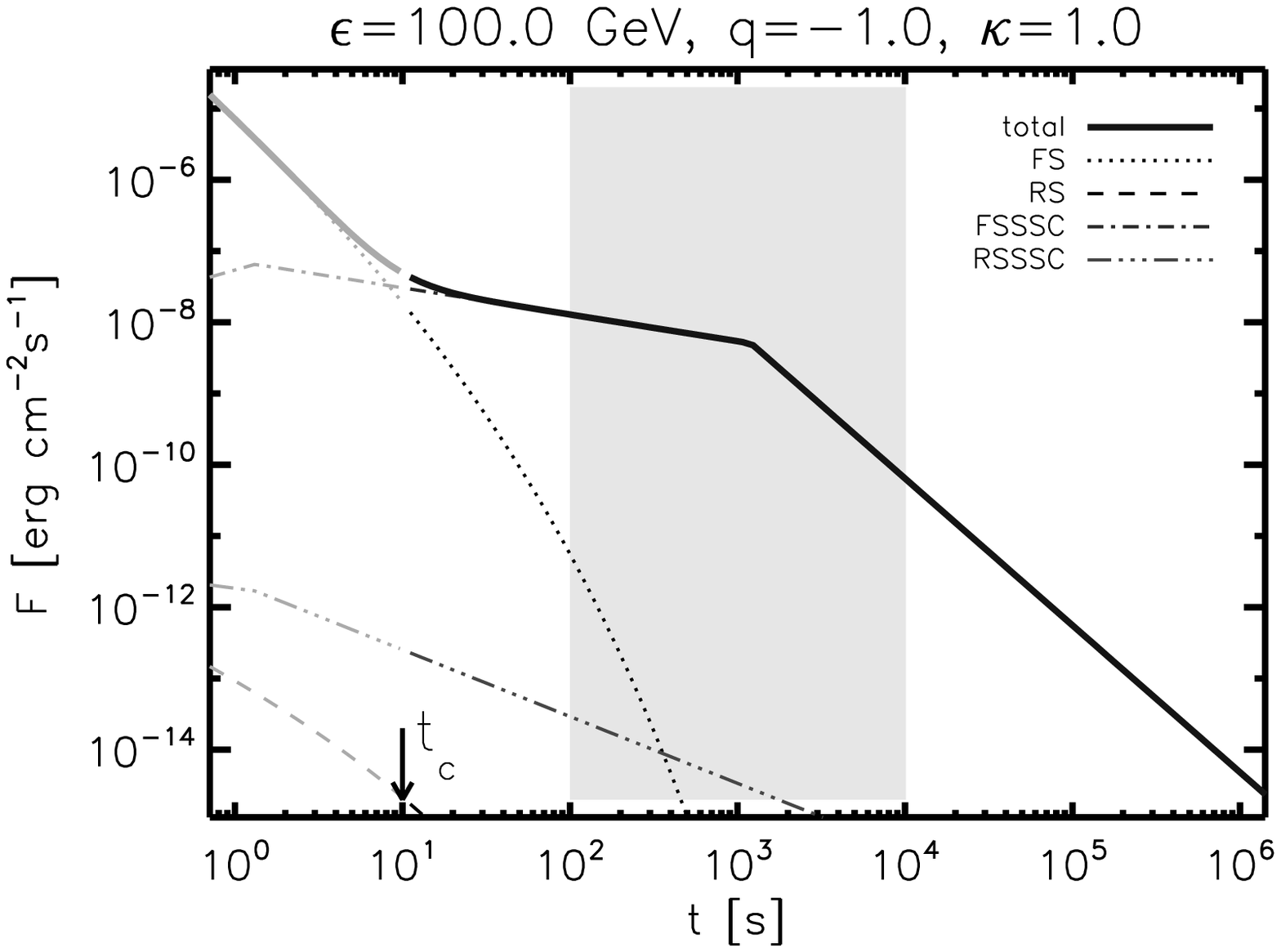}
\includegraphics[width=0.899\columnwidth,angle=0]{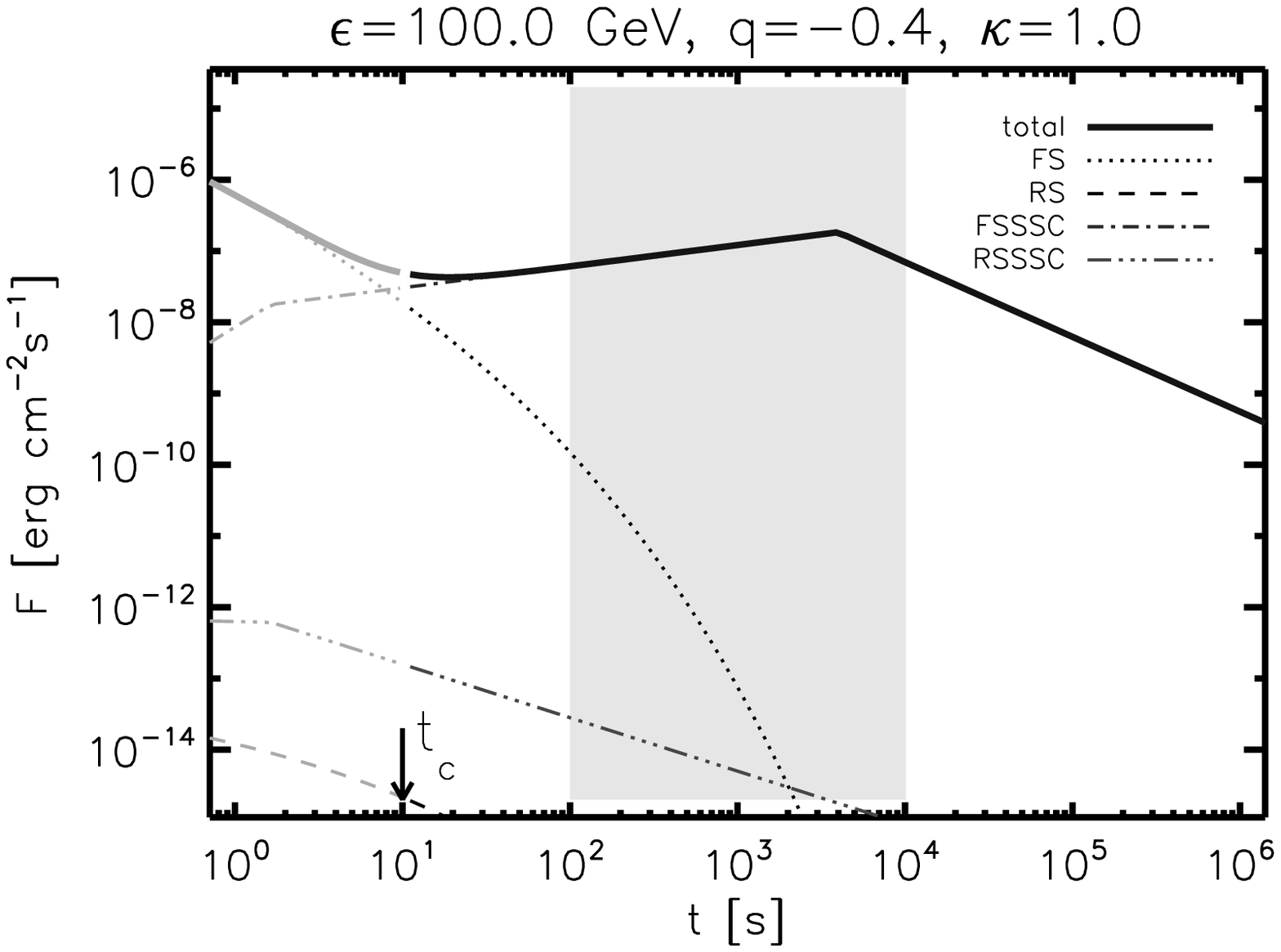}
\caption{Lightcurves in the continued energy injection case at $100 \GeV$.  The
$q$ parameter is $q=-1$ (top) and $q=-0.4$ (bottom).  Here $t_c$ marks the time
when the initial injection term is equal to the continuously injected energy.
Other notation is similar to Figure 1.} \label{fig:tempinj} 
\end{center}
\end{figure}

The continuous injection model lightcurves, based on the discussion in \S
\ref{sec:continj}, are qualitatively similar to those for the refreshed shock
cases.  For comparison with the refreshed shock scenario, we use the same
parameters as in Section \ref{sec:refresh}.  The parameter  $t_0$ is the start
of the self similar phase. The value of $t_c$ in figure \ref{fig:tempinj} is
defined as either the time  when the two terms are equal on the right hand side
of equation \ref{eq:e}, or the time $t_0$, whichever is larger.  This ensures
that the lightcurve is in the self-similar phase and dominated by the
continuous injection.

The continuous late injection phase can yield reverse shock and SSC components
similarly to the case of refreshed shocks.  For simple cases we find a
one-to-one correspondence between $s$ and $q$, e.g.  in the homogeneous ($g=0$)
ISM case, $q=2(s-5)/(s+7)$.  To illustrate the similarity, we calculated sample
lightcurves based on \citet{Zhang+01mag} (see Figure \ref{fig:tempinj}).

\section{Prospects for Observational Detection}
\label{sec:det}

The next generation Cherenkov telescopes may be advantageous for detecting  the
electromagnetic counterpart of a gravitational wave event
\citep{Bartos+13gwcta}.  We concentrate on the afterglow detection. Pointing
instruments generally require a time delay from the satellite trigger time to
the start of the GeV-TeV observations.  Observatories with extended sky
coverage can in principle detect both the prompt and the afterglow emission,
though with a lower sensitivity.

\begin{figure}[htbp]
\begin{center}
\includegraphics[width=0.899\columnwidth,angle=0]{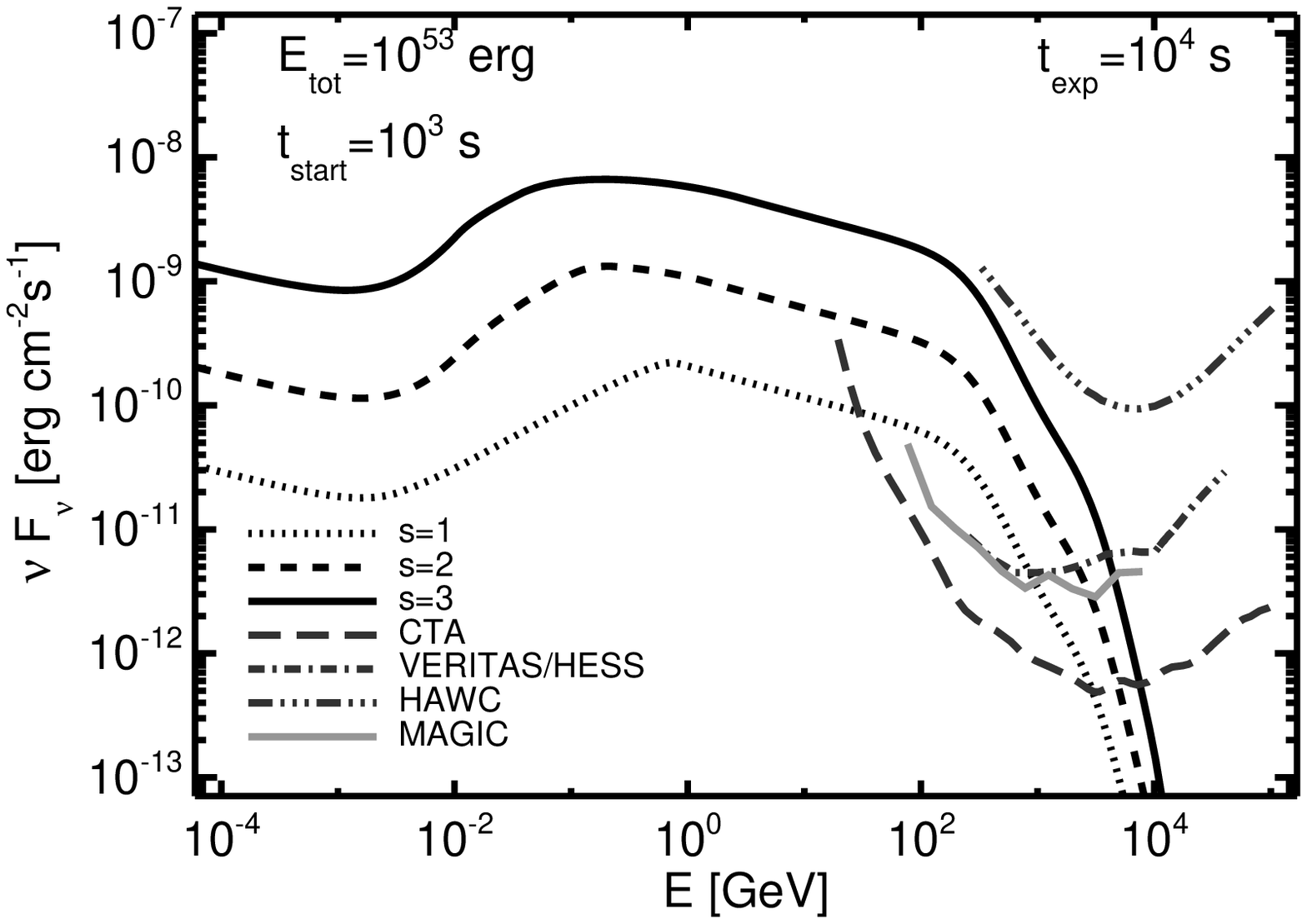} 
\includegraphics[width=0.899\columnwidth,angle=0]{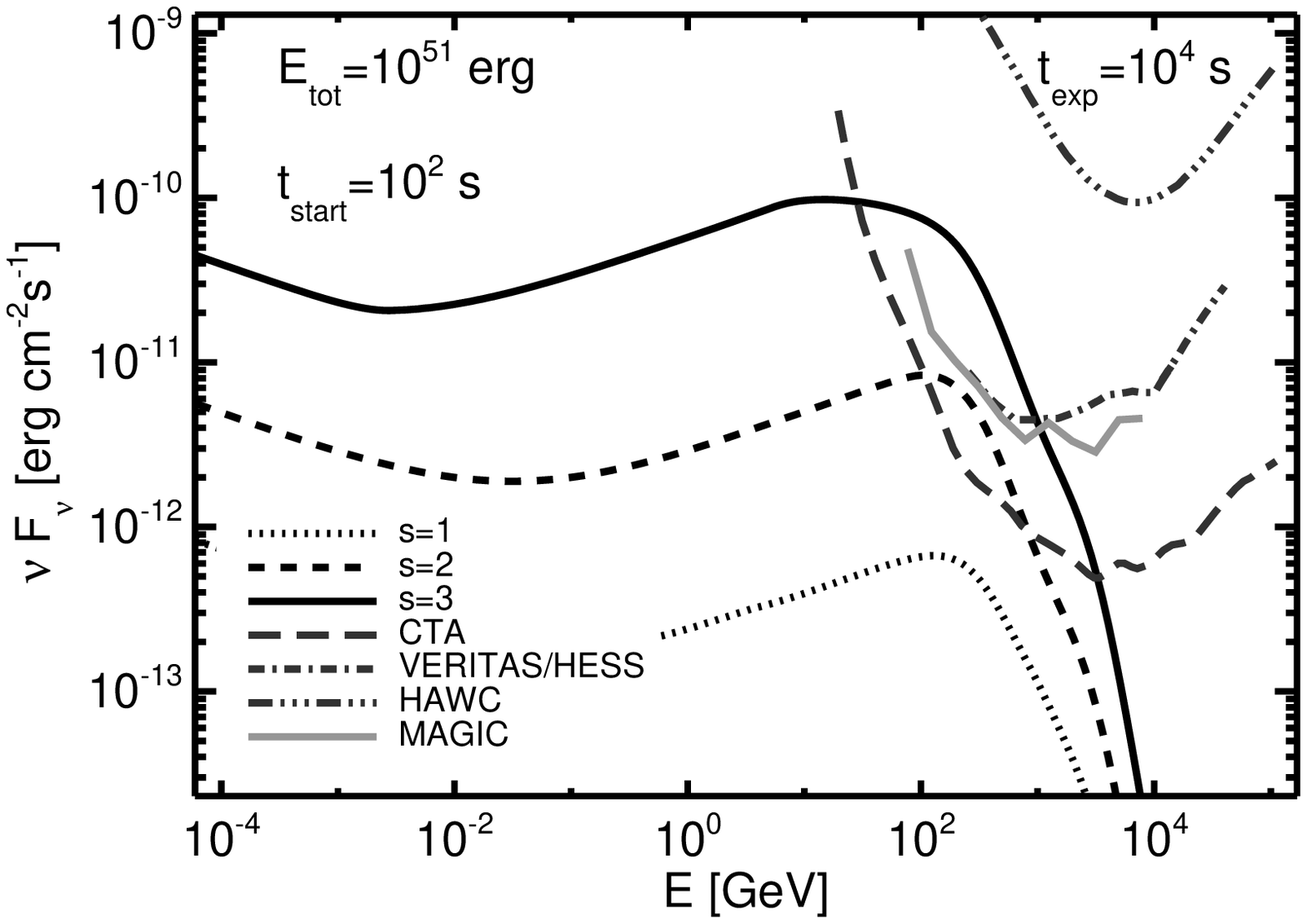} 
\caption{Average spectrum for the $s=\{1,2,3\}$ cases. The upper panel shows
the example case presented in Section \ref{sec:lc} for observations starting at
$t_{\rm start}=10^3 \s$ and lasting for $dt = 10^4 \s$. The lower figure is a
model with total energy of $E_{\rm t}=10^{51} \erg$ and  $t_{\rm start}=10^2
\s$, $dt = 10^4 \s$.  Overlaid are the differential sensitivity curves of CTA,
VERITAS/HESS, MAGIC and HAWC.}
\label{fig:spec} 
\end{center}
\end{figure}

\subsection{Detectability with current and future telescopes}
\label{sec:tel}
{\it CTA - } We calculate the sensitivity of CTA following the treatment of
\citet{Actis+11CTA}, based on their Fig. 24. The detection capabilities of the
future CTA telescope in the context of GRBs was discussed at length by
\citet{Inoue+13GRBCTA}.

{\it VERITAS and HESS - } These instruments have similar sensitivities for our
purposes. VERITAS can slew on the order of $100 \s$ to a source with favorable
sky position. For a compilation of sensitivity curves see \citet{Ab+13HAWCsens}
and references therein, for HESS see \citet{Aharonian+04crab}.

{
{\it MAGIC - } This instrument has the best sensitivity at  $\sim 100\GeV$
\citep{Aleksic+12MAGIC}. It has comparable sensitivity to VERITAS/HESS up to
$\sim 100 \TeV$ (see Figure \ref{fig:spec}). Furthermore, the average slewing
time for MAGIC is $\sim 20 \s$ after the alert, which makes it well suited for
early follow-up \citep{Garczarczyk+09MAGIC}.  }

{\it HAWC - } The HAWC instrument is a water Cherenkov observatory with large
field of view and duty cycle. It can potentially detect both the prompt and the
afterglow emission. Here we take the HAWC sensitivity for steady sources, which
is a crude approximation for the long-lasting lightcurves of our models,
assuming an observation lasting $10^4\s$ \citep{Ab+13HAWCsens}.

{\it Detection - } In general terms, models with energy injection due to either
refreshed shocks or continued outflow result in a larger radiation flux at late
times than the conventional models. This is evident from Figure \ref{fig:spec}.
The interplay of the brightness and distance of the burst and the slew time of
the telescope determines the detection.  We assume an average time delay after
the satellite trigger time of $t_{start}=10^3$ s.  In the upper panel the total
energy is $E_t=10^{53}$ erg and the exposure time is $10^4 \s$, while in the
lower panel of the figure, we reduced the total energy budget of the GRB to
$E_t= 10^{51} \erg$ for the same exposure time.  Except for the no-injection
(no LF spread, $s=1$) model in the $E_{\rm t}=10^{51}\erg$ case, all of the
models can be detected at least marginally with CTA and some with VERITAS{ ,
MAGIC and HESS}.  The HAWC instrument can only marginally detect afterglow
lightcurves in the most optimistic cases presented here. HAWC is better suited
for observing the prompt phase, for which there is a more sensitive data
analysis in place, but may be able to detect the very early afterglow.

These figures show that for standard afterglow parameters and average response
time, observing GeV-TeV range range radiation from short GRB afterglows is
promising with current instruments. 

\subsection{High energy temporal indices}
\begin{figure}[htbp]
\begin{center}
\includegraphics[width=0.899\columnwidth,angle=0]{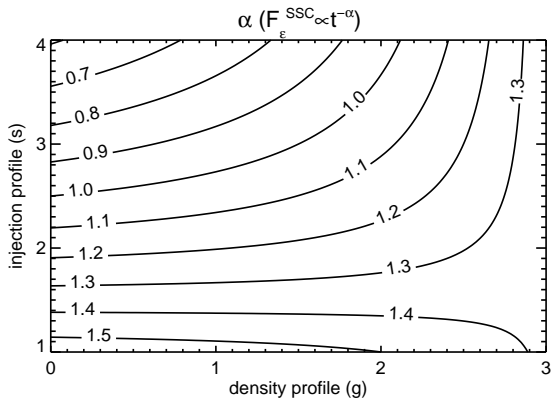}
\caption{Contour map of the temporal index as a function of the injection
($s$)- and circumstellar density ($g$) profile (see Equation \ref{eq:tidx}).
These are the  temporal slopes expected at late times for observing energies
$\ve_{\rm obs}>\max\{\ve_m^{\rm FSSSC},\ve_c^{\rm FSSSC}  \}$. Here, $p=2.5$
and possible Klein-Nishina or EBL effects are not taken into account. These
would result in a higher $\alpha$.}
\label{fig:sg} 
\end{center}
\end{figure}

Both cooling and injection SSC energies are a decreasing functions of time. At
$\sim 100 \GeV$, observations will occur above the highest break frequency and 
the temporal slope can be calculated from (all SSC parameters): 
\begin{eqnarray}
F_\ve&=&F_{\rm peak}
(\ve_c/\ve_m)^{-(p-1)/2} (\ve_{\rm obs}/\ve_c)^{-p/2}\\
&\propto& t^{-( -32 -8s + 10g + 2gs + p(36-11g+gs))/4(7+s-2g)}.\label{eq:tidx}
\end{eqnarray}
The slope will be the same irrespective of the cooling regime. This can be compared 
with measurements, in order to constrain the values of $s$. For example if $g$ and
$p$ are known from lower energy observations, the measured high energy slope is
very sensitive to the value of $s$.  Figure \ref{fig:sg} shows an example of
slopes for the forward shock SSC at late times for $p=2.5$.

{ 
\subsection{X-ray counterparts}\label{sec:x}
As mentioned in the introduction, in the case of gravitational waves,
electromagnetic follow-up observations are crucial. That is also the case in
the event of a detection of a GRB at TeV energies.  Currently the Swift XRT
\citep{burrows05} is most likely to provide X-ray follow-up on timescales of
$100 s$ after the trigger \citep{Evans+12GWfollowup}.  Similarly to the GeV-TeV
range, we have calculated the lightcurves at $10 \keV$ (see Figure
\ref{fig:lcX}).  The X-ray afterglow is dominated by synchrotron radiation. The
temporal slope for $s>1$ is visibly shallower than the $s=1$ case, and that is
the reason this model is favored for the interpretation of the X-ray plateau.
At late times ($10^5-10^6 \s$), the FSSSC provides a bump in the lightcurve.
This late after the trigger only the brighter bursts are detected with XRT, and
typically no unambiguous bump is observed \citep{Zhang+14chandra}.

In contrast to the X-rays, the GeV-TeV range lightcurves are dominated by the
SSC from the FS. In the framework of the refreshed shock model, the
difference between the temporal slopes can be expressed as (from
\citet{Sari+00refresh} and Equation \ref{eq:tidx}): $\Delta \alpha =
\alpha_X-\alpha_{\rm TeV} = (3-g)(p-2)/(7-g+s)$, where the X-ray has the
shallower decay. For e.g. in case of $p=2.5$ the X-ray slope is flatter at most
by $\sim 0.2$, where the expression for $\Delta \alpha$ has its maximum, for
$s=1$ and $g=0$ (no injection and constant ISM case). In general,  $\Delta
\alpha \lesssim 3(p-2)/8$.

\begin{figure}[htbp]
\begin{center}
\includegraphics[width=0.899\columnwidth,angle=0]{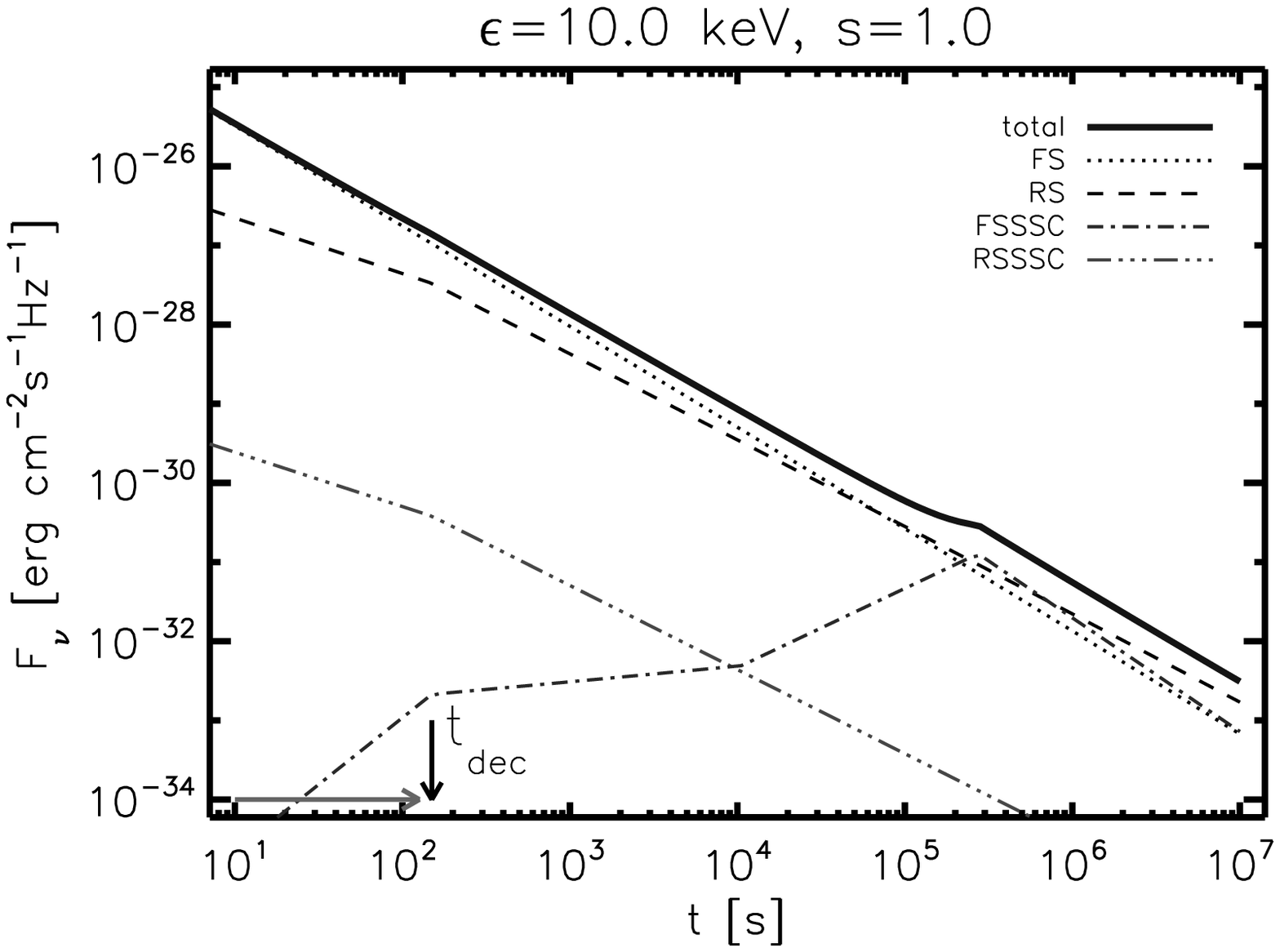}
\includegraphics[width=0.899\columnwidth,angle=0]{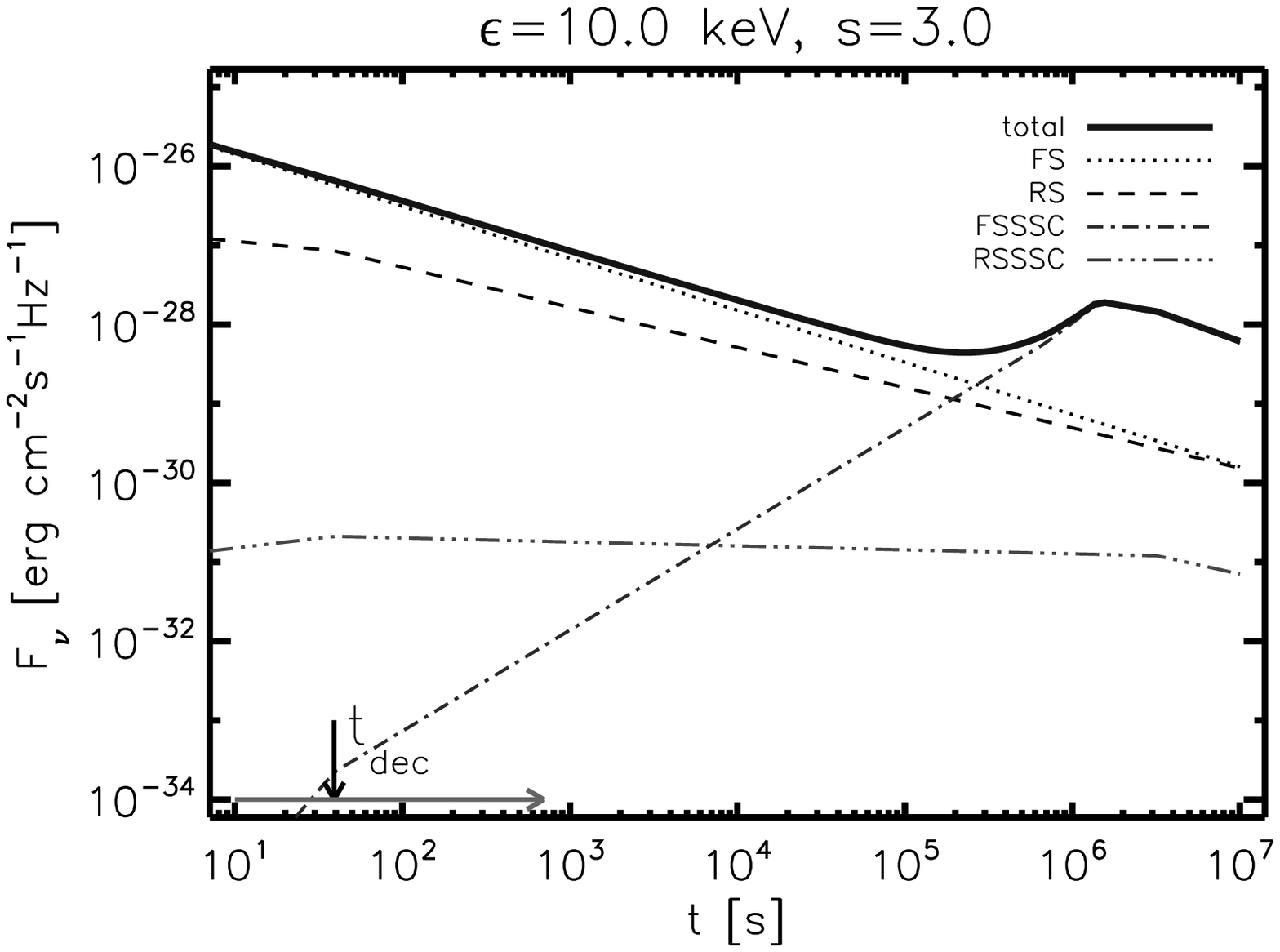}
\caption{X-ray lightcurve for an afterglow with standard parameters. } 
\label{fig:lcX} 
\end{center}
\end{figure}

}

{ 
\subsection{The case of GRB 090510}

The high-energy lightcurve of the  archetypical short  GRB 090510 is consistent
with a single power law.  Certainly, up to now there is no definite proof of an
SSC component at high energies. The GeV range lightcurves can be explained by
one component (see e.g. \citet{DePasquale+09-090510ag} or \citet{Kouv+13nustar}
for GRB 130427A). As shown in Figure \ref{fig:lc1}, the SSC component starts to
dominate at a later time and there is a pronounced "bump" feature in the
lightcurve. This feature becomes more prominent with increasing $s$.  There are
two possibilities how the SSC could be dominant in the GRB afterglows.  In the
first case, the whole GeV-range flux is attributed to SSC. This needs fine
tuned parameters as the SSC peak needs to be very early.  The second option is
to have a very smooth transition in time between synchrotron dominated and SSC
dominated times. This requires less fine tuning because  e.g. the temporal
slopes of the two components will not be too different (see section
\ref{sec:x}) but the fluxes have to match within errors. Thus we speculate that
a bump feature is present in the lightcurve (see Fig 1. in
\citet{DePasquale+09-090510ag}), but not significant. This points to a low
value of $s$, close to $1$}

\section{Discussion and Conclusions}
\label{sec:disc}
Thus far there is a dearth of short GRB afterglow lightcurves observed above
$\GeV$ energies, compared to long GRBs. Nonetheless, future and current
observatories offer realistic prospects for detecting GeV-TeV range emission
from these sources, which are also the prime targets for gravitational wave
detectors. 

We have calculated the TeV range radiation for the refreshed shock and the
continued injection afterglow models of short GRBs. The refreshed shock model
involves a range of injected LFs, the slower material catching up at later
times, reenergizing the shocks, and qualitatively similar results are obtained
in continuous injection models.

Our aim was to show that even though many GeV lightcurves decline rapidly, the
sensitivity of current and future TeV range instruments is sufficient to
detect short GRB afterglows in the framework of the refreshed shock model,
where the decline of the flux can be slower. In the relevant GeV-TeV range we 
found that the afterglow SSC components are the main contributors to the flux.  
The usual time dependences for the non-injection cases are recovered by setting 
$s=1$ and $g=0$ or $g=2$.

We also showed that in the simple case of an ISM environment and adiabatic
afterglow, there is a one-to-one correspondence between the continuous
injection model and the instantaneous injection refreshed shock model with a
range of LFs. A possible distinction between these scenarios can be made by
following the X-ray lightcurve. In the continuous injection model the plateau
is expected to end abruptly \citep{Zhang13frb}. 

We have calculated the dependence of the peak time (peak or plateau in the
lightcurve) on the microphysical parameters and  found that it correlates
positively with the $\epsilon_e$, $\epsilon_e$ and $g(p)$ parameters, while it
anticorrelates with the external density $n_{\rm ext}$.  For late times and
high energies we have calculated the expected temporal slope. These can be
used, when the first TeV range measurements are carried out, to constrain the
$s$ index of the LF distribution.

The general behavior of a model with a Lorentz factor spread or late injection
is that for the same total energy output it starts out dimmer and has a flatter
decline at later times.  It has a more pronounced plateau phase for higher $s$
parameter, ending at $\sim 1000\s$.  The importance of the refreshed shock and
continuous injection models lie in supplying energy to the external shock
complex at late times, flattening  the afterglow decay, and thus making it more
favorable for detection at TeV energies.  The flattening in the lightcurve is
most notable at $10^2-10^3 \s$ after the burst trigger. Our calculations show
that current and next generation ground-based instruments will be well suited
to detect the GeV-TeV emission of short bursts with extended emission. In the
event of a gravitational wave trigger, even if Swift or Fermi are in Earth
occultation, the ground-based GeV-TeV detections could provide a good
counterpart localization.

We thank NASA NNX13AH50G and OTKA K077795 for partial support, and { David
Burrows, Charles Dermer,} Abe Falcone { and}  Dmitry Zaborov   for
discussions. { We further thank Stefano Covino and Markus Garczarczyk for
providing the MAGIC sensitivity curve and the referee for a thorough report.}

\bibliographystyle{hapj}

\end{document}